\documentclass[nofootinbib,a4paper,aps,pra,showpacs,preprintximbers,twocolumn]{revtex4-2}
\usepackage{cmap}
\usepackage{graphics,graphicx,epsfig}
\usepackage{epstopdf}
\usepackage[centertags]{amsmath}
\usepackage{amsfonts}
\usepackage{amssymb}
\usepackage[cp1251]{inputenc}
\usepackage[english]{babel}
\usepackage{graphicx}
\usepackage{amsthm,color}
\usepackage{euscript}
\usepackage{indentfirst}
\usepackage{xcolor}
\usepackage{braket}
\usepackage{type1ec}%
\usepackage{physics}
\usepackage{tabularx}
\usepackage[T1]{fontenc}	
\usepackage{enumitem}
\DeclareGraphicsExtensions{.pdf,.png,.jpg,.eps}

\newlist{steps}{enumerate}{1}
\setlist[steps, 1]{label = Step \arabic*:}
\begin{document}
	\def\BY{\begin{eqnarray}}
		\def\EY{\end{eqnarray}}
	\def\L{\label}
	\def\nn{\nonumber}
	\def\ds{\displaystyle}
	\def\({\left (}
	\def\){\right )}
	\def\[{\left [}
	\def\]{\right]}
	\def\<{\langle}
	\def\>{\rangle}
	\def\h{\hat}
	\def\td{\tilde}
	\def\r{\vec{r}}
	\def\ro{\vec{\rho}}
	\def\h{\hat}
	\title{Parallel two-qubit entangling gates via QND interaction controlled by rotation}
	\author{E.A. Vashukevich$^1$}
	\author{T. Yu. Golubeva$^{1,2}$}
	
	\affiliation{$^1$Saint-Petersburg State University, Universitetskaya Nab. 7/9, St. Petersburg 199034, Russia,\\ $^2$ Laboratory of Quantum Engineering of Light, South Ural State University, pr. Lenina 76, Chelyabinsk, 454080, Russia}
	\begin{abstract}
The paper presents an analysis of entangling and non-local operations in a quantum non-demolition (QND) interaction between multimode light with orbital angular momentum and an atomic ensemble. A protocol consists of two QND operations with rotations of quadratures of atomic spin coherence and light between them. This protocol provides a wide range of two-qubit operations, while the multimode nature of the chosen degrees of freedom allows the implementation of parallel operations over multiple two-qubit systems. We have used the formalism of equivalence classes and local invariants to evaluate the properties of two-qubit transformations.  It is shown that, when selecting suitable values of the governing parameters, such as the duration of each of the two QND interactions and the rotation angles of the qubits, the protocol allows to realise a deterministic non-local SWAP operation and entangling $\sqrt{SWAP}$ operation with probability 1/3.
	\end{abstract}
	\maketitle
	\section{Introduction} 
Quantum computation with qubits has been deeply investigated theoretically \cite{QCQ2,QCQ2} and has numerous experimental realizations \cite{QCE1, QCE2,QCE3, Lukin1}. Today, one of the most important challenges for the scientific community is to increase the number of physical qubits, which could be entangled. It is necessary both for achieving confident quantum supremacy and for building quantum error correction codes \cite{QErr} using redundant encoding.

Light with orbital angular momentum (OAM) seems to be a promising object for application in the quantum informatics problems, since OAM can take any integer values \cite{allen}, which can be used for cryptography protocols \cite{crypt1, crypt2}, construction of qudits \cite{qudits, Goy, qudits2}, and also turns out to be useful in the problems of quantum communication \cite{comm1, comm2}. One of the advantages of using light as a physical platform for quantum applications is the possibility of organizing the interaction of light with an atomic ensemble, for example, for entangling spatially separated atomic ensembles \cite{entAtom1, entAtom2} and for storing quantum fields in a quantum memory \cite{QM1, QM2, QM3}.  It has been demonstrated \cite{Convers} that considering interaction of OAM light with an atomic ensemble it is possible to identify OAM modes of collective spin coherence of the ensemble, pairwise interacting with modes of the field. One could control the interaction regimes and efficiency by switching the OAM of the driving field, as well as by changing the scheme geometry. Various quantum memory protocols for light with OAM \cite{QMOAM1, QMOAM2}, as well as realizations of single-qudit \cite{qudits, zail, qudits1} operations are demonstrated.    

To construct a universal set of operations in discrete variables \cite{univer}, the construction of two-mode gates is required. In \cite{BVG}, we have examined an atomic-field system with quantum non-demolition (QND) interactions between field modes with a certain OAM and collective spin coherence modes.  We have demonstrated ways to encode the logical states of multiple qubits on physical atomic and field states for different spatial profiles of the driving fields. It has been shown that it is possible to identify many independent two-qubit subsystems. We have considered a protocol consisting of QND operation, rotation of oscillators' quadratures and another QND operation, and identified a set of parameters that allows us to perform a parallel deterministic n-two-qubit SWAP ("n-" denotes the total number of two-qubit subsystems, which we can operate with) operation over the entire set of two-qubit subsystems. But, despite the importance of the SWAP operation for constructing quantum algorithms \cite{SWAP1, SWAP2}, this operation is not entangling.

In this paper, we focus on the entangling properties of the QND-rotation-QND transformation, expecting to reveal ways to perform n-two-qubit entangling operations in parallel, which can significantly reduce computational time \cite{parallel}, and can also be used in error-correcting codes for codewords' creation and manipulation \cite{QErr2}. In order to distinguish gates only by entangling properties, ignoring local single qubit rotations, we use the formalism of equivalence classes and local invariants described in \cite{makhlin}. The paper discusses in detail the restrictions on the equivalence classes imposed by the nature of the considered interaction, and investigates the relation of the qubit rotation angles and the effective  interaction constants to the equivalence class of the two-qubit transformation.  The paper also presents an analysis of the probabilities of entangling gates and indicates the values of governing parameters corresponding to the highest probabilities. 
	\section{Model and definitions}
\begin{figure}
	\includegraphics[width=8.6cm]{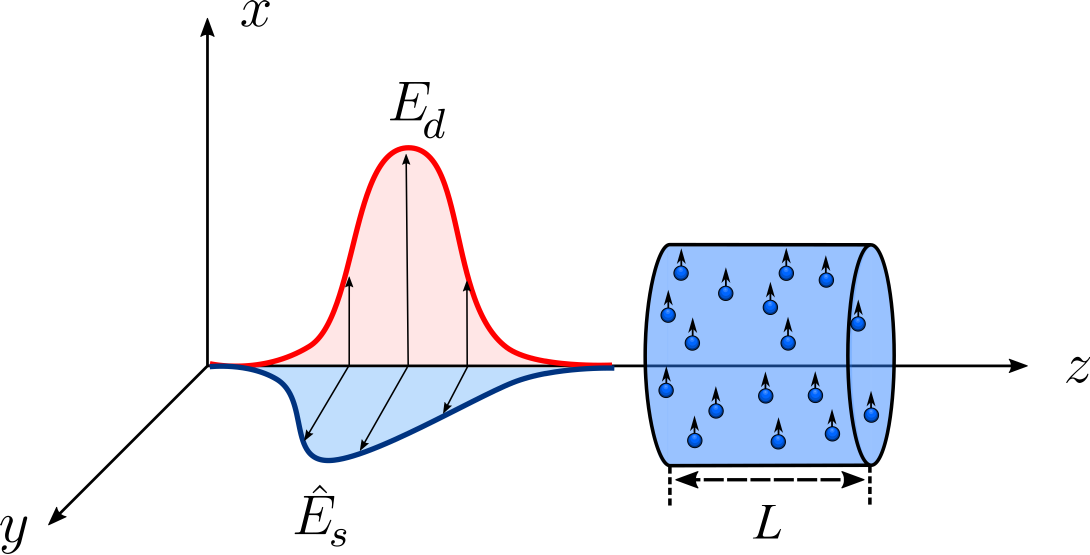}\\ \includegraphics[width=8.6cm]{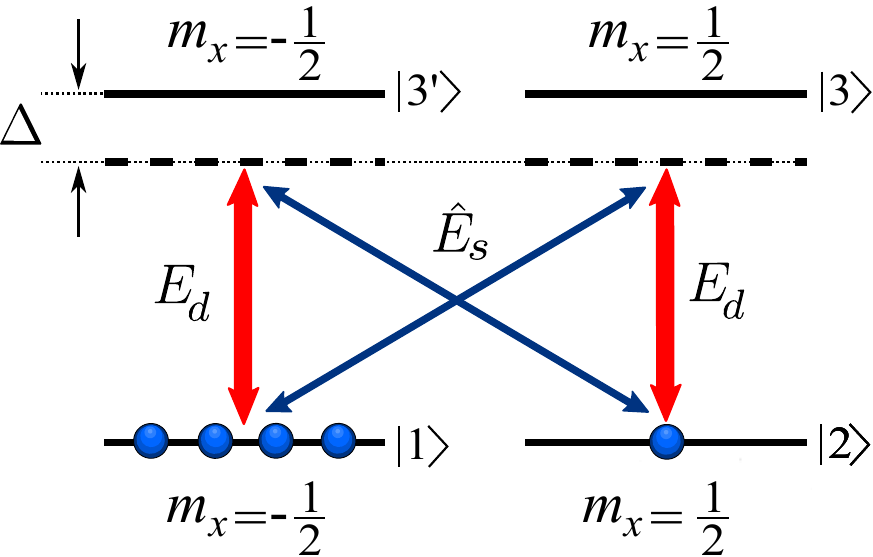}
	\caption{Schematic representation of the interaction geometry and the atomic energy levels.  Here $\Delta$ is the detuning of light frequency from atomic transitions, classical and quantum fields have polarization $\vec{e}_{x}$ and $\vec{e}_{y}$, respectively, so that the driving field acts on the transitions $\ket{1}-\ket{3}^{\prime}$ and $\ket{2}-\ket{3}$ and the quantum field acts on the transitions $\ket{1}-\ket{3}$ and $\ket{2}-\ket{3}^{\prime}$. Initially, an atomic ensemble prepared in the state $\ket{1}$ with the average collective spin directed along the axis $x$ and the magnetic spin momentum $m_x=-\frac{1}{2}$.}\label{FigSys}
\end{figure}

First of all, we want to discuss the physical model of the proposed protocol. Following \cite{BVG}, we consider the Faraday interaction of a multimode quantum field, which is a superposition of spatial modes with different values of the orbital angular momentum, with an ensemble of cold four-level atoms (see Fig. \ref{FigSys}). The interaction is carried out by a multimode driving field with an OAM, which we characterize by a set of Rabi frequencies $\Omega_n(z,t)=\frac{E_{n}(\vec{r},t)d_{23}}{\hbar}$, where $E_{n}$ is the amplitude of the driving field mode with OAM value $n$, $d_{23}$--the dipole moment of the transition. 
It was shown that in the atomic ensemble one can identify a set of spin coherence modes with OAM and introduce bosonic operators $\h b_l$ characterizing the annihilation of excitations in the modes, where the index $l$ indicating the OAM of the modes. As a result, the light-atomic interaction can be effectively described by the Hamiltonian:
\BY
&&\hat{H}_{I}=-\frac{\sqrt{2}\hbar g \sqrt{N}}{\Delta \sqrt{c}}\int dz\;\sum\limits_{m,l} \left(\chi_{l-m}\Omega_{l-m}\left[\hat b_{l}^\dag\hat{a}_{m}+ \h b_{l}\hat{a}^{\dag}_{m}\right]\right.\nonumber\\
&&\left.-\chi_{l+m}\Omega_{l+m}\left[\hat b_{l}^{\dag}\hat{a}_{m}^{\dag}+\hat b_{l}\hat{a}_{m}\right]\right), \label{hqnd}
\EY
where $\h a_n$ is the photon annihilation operator in the field mode with index $n$, $g$ -- atom-field interaction constant, $N$ is the average density of atoms in the ensemble, $\Delta$ is the detuning of the quantum field frequency from the atomic transition, $\chi_{l-m}$ -- the overlapping integral of the spatial profiles of the quantum field mode with index $l$, the atomic spin coherence mode with index $m$, and the driving field mode  with index $l-m$, normalized by the area of the beam waist. The presented Hamiltonian consists of two parts -- the beam splitter Hamiltonian (terms proportional to $\h a_l^\dag\h b_m$ and $\h a_l\h b_m^\dag$) and the parametric gain Hamiltonian ($\h a_l^\dag\h b^\dag_m$ and $\h a_l\h b_m$). In our previous paper, we considered several interaction regimes determined by the value of the OAM of the driving field, but within the framework of this paper we  restrict ourselves to describing only the case of a single driving field mode with OAM equal to 0. In this case, the modes with the identical indices are interacting. The Hamiltonian (\ref{hqnd}) describes the pairwise interaction of modes, and one can write the solutions of Heisenberg equations for field and atomic operators in the form of the Bogoliubov transformation with the effective interaction constant $\xi$:
\BY
&& \begin{pmatrix}\h A_m^{\dag}\\\h B_m^{\dag}\end{pmatrix}^{out}=-\frac{i}{2}\[G\begin{pmatrix}
	\h A_m^{\dag}\\\h B_m^{\dag}\end{pmatrix}^{in}+ R\begin{pmatrix}
	\h A_m\\\h B_m\end{pmatrix}^{in}\],\\
 &&G=\begin{pmatrix} 2i &\xi  \\
	\xi &2i\end{pmatrix} ;\;\;R=\begin{pmatrix} 0 &-\xi\\
	-\xi &0\end{pmatrix};\\
	&&\xi=2\sqrt{\frac{2 N}{c}} g\chi_{m-k}\sqrt{\int\limits_0^T\frac{\Omega^{2}_{0}(0,t)}{\Delta^2}dt}\label{xi}.
	\EY
Here $T$ is the interaction time chosen in such a way that $cT\gg L$. Varying the pulse duration of the driving field, as well as the pulse energy, allows us to change the effective interaction constant $\xi$. The operators $\h A_m, \h B_m$ are defined as follows:
\BY
&&\hat{A}_{m}(z)=\frac{\int\limits_0^T \Omega_{0}(t)\[\hat{a}_{m}(z,t)+\hat{a}_{-m}(z,t)\] dt}{\sqrt{2\int\limits_0^T \Omega_{0}^{2}(t)dt}},\nn\\
&&\hat{B}_{m}(t)=\frac{1}{2\sqrt{L}}\int\limits_0^L \;\[\hat{b}_{l}(z,t)+\hat{b}_{-l}(z,t)\]dz.\label{QM}
\EY
In the expressions for operators $\hat{A}_{m}(z)$ and $\hat{B}_{m}(t)$, the sums of field ($\h a_l$) and atomic ($\h b_l$) operators with opposite indices appear, i.e. $\hat{A}_{m}(z),\hat{B}_{m}(t)$ are annihilation operators in the superpositional OAM modes. For each pair of such operators with index $m$ the evolution of the Hamiltonian (\ref{hqnd}) reduces to the atomic-field QND interaction.

Since our goal is to describe the entangling interaction between two qubits, following the  \cite{muzhik,polzik}, we consider a multipass scheme consisting of a QND interaction with constant $\xi_1$, a further rotation of the light and atomic quadratures by angles $\theta_1, \theta_2$, respectively, and another QND with constant $\xi_2$. Using similar protocol in continuous variables allows to realise QND quantum memory or two-mode squeezing, depending on the rotation angles and values of the constants. However, our approach is multimode, so we simultaneously control the evolution of a whole set of modes. We also consider the evolution of the system in terms of qubit transformations, and the multimode nature of the protocol means that it is possible to perform transformations for multiple two-qubit subsystems in parallel. 

The physical implementation of the necessary manipulations with light and atomic systems is quite intuitive: the rotation by an arbitrary angle $\theta_1$ in the light system can be described as an additional phase of the light beam, which can be added by using phase plates or a delay line, the rotation of atomic variables can be performed by applying an external magnetic field along the $x$-axis, which lead to the rotation of a small spin component orthogonal to the $x$-axis in the $yOz$-plane (see Fig. 1).  The result of the evolution of operators in the QND-rotation-QND (QRQ) protocol can be written in the form of the Bogoliubov transformation:
\BY
&&\begin{pmatrix}\h A_m^{\dag}\\\h B_m^{\dag}\end{pmatrix}^{in}=-\frac{i}{2}\left[\td{G}
\begin{pmatrix}  \h A_m^{\dag}\\\h B_m^{\dag}\end{pmatrix}^{out}+\td{R}
\begin{pmatrix}  \h A_m\\\h B_m\end{pmatrix}^{out}\right],\label{Bog}\\
&&\td{G}=\begin{pmatrix}2i e^{i\theta_1}-\xi_2\xi_1\sin{\theta_2}&\;\xi_2e^{-i\theta_1}+
	\xi_1e^{-i\theta_2} \\\xi_2e^{i\theta_2}+
	\xi_1e^{i\theta_1} &\;2i e^{i\theta_2}-\xi_2\xi_1\sin{\theta_1}
\end{pmatrix},\hspace{15pt}\\
&&\td{R}=-\begin{pmatrix}-\xi_2\xi_1\sin{\theta_2}&\;\xi_2e^{-i\theta_1}+
	\xi_1e^{i\theta_2} \\\xi_2e^{-i\theta_2}+
	\xi_1e^{i\theta_1} &\;-\xi_2\xi_1\sin{\theta_1}
\end{pmatrix},\hspace{11pt}
\EY
where the constants $\xi_1$ and $\xi_2$ account for the two stages of QND interactions and are defined by the Eq. (\ref{xi}).

 Before describing the possible two-qubit transformations, we need to define the basis states of the logical two-qubit space through the physical states of field and atomic modes. In \cite{BVG} it have been shown that such a basis can be defined through the Fock states of atomic and field modes with single excitation:
\BY
\ket{1}_{m,L} \equiv\h A_m^{\dag}\ket{0}_{m,L};\;\;\ket{1}_{k,A}\equiv\h B_k^{\dag}\ket{0}_{k,A}.\;\;
\EY
Here $\ket{0}_{m,L},\ket{0}_{k,A}$ are the vacuum states of the field and atomic systems, respectively, the indices $m,k$ are related to the value of the OAM; the upper index $in$ is omitted. By restricting the maximum OAM of quantum states by some number $K-1$,  on the set of physical states we can identify $K/2$ independent two-qubit subsystems, where the basis logic states of qubits are defined as follows:
\BY
&&\ket{0}^j_1\equiv\ket{1}_{2(j-1),L} ;\;\;\ket{1}^j_1\equiv\ket{1}_{2j-1,L}; \\
&&\ket{0}^j_2\equiv\ket{1}_{2(j-1),A} ;\;\;\ket{1}^j_2\equiv\ket{1}_{2j-1,A} .
\EY

The index $j\in[1,K/2]$ refers to the number of the two-qubit subsystem, the lower indices 1 or 2 numbering the qubits within the same subsystem. Such logical states of qubits are encoded by physical states with single excitations in field or spin atomic coherence modes. Thus, each two-qubit system contains two excitations: one in the superposition state of two light modes, and another one -- in the superposition state of two spin coherence modes. The initiation of a system of $K/2$ two-qubit subsystems requires $K$ excitations.  The Bogoliubov transformation (\ref{Bog}) is carried out for all pairs of atomic and field operators with different indices simultaneously, which in terms of qubit states' evolution means parallel evolution of all two-qubit subsystems. Moreover, all effective interaction constants can also be equal to each other, which is achieved by geometrical displacement of driving and quantum beams \cite{Convers}.

The challenge of performing two-qubit entangling gates in our case from the perspective of the model is the problem of organising QND interaction between modes with an certain effective constant. The interaction scheme presented in Fig. 1 is similar to QND quantum memory schemes, and in many aspects the constraints on the cooling depth of the atomic ensemble, trap parameters and pulse shapes of the driving and quantum fields in the considered case are similar to those in  quantum memory schemes \cite{QMOAM1, QMOAM2}. An important difference, at first, is the requirement on the length of the atomic trap in comparison to the Rayleigh range of the driving beam, as well as the ratio of the transverse area of the ensemble to the beam width, which are described in detail in [20]. Secondly, if we are not interested in the problem of storing information for a long time, the implementation of gates is obviously faster than the quantum memory protocol due to the lack of storage time -- we only need to perform one memory step, called the write-in step, which is usually much shorter than the storing time. This also gives better fidelity values because the decoherence of atomic states due to atomic motion will be smaller due to the shorter protocol execution time, which weakens the requirement for atomic temperature. At the same time, it seems logical to combine quantum memory and the execution of operations on quantum states -- in this case the physical parameters will be limited primarily by the required storage time. Further we will speak only about distinct, additional requirements to interaction constants for performing specific nonlocal and entangling operations.

The considered transformation of atomic and light states depends on four parameters, namely the two real positive constants of the first and second QND interactions, $\xi_1$ and $\xi_2$, which are determined by the atomic density and the effective integral time of the interaction (see (\ref{xi})), and the two rotation angles $\theta_1,\theta_2$. To investigate the entangling properties of the two-qubit transformation, it is useful  to consider the technique of local invariants \cite{makhlin}, briefly described in the next section.
\section{Local invariants and equivalence classes}
\subsection{Brief review of local invariants}
A local unitary two-qubit transformation is a transformation $\mathcal{U}$ such that $\mathcal{U}\in SU(2)\otimes SU(2)$, that is, a transformation which reduces to a direct product of independent single-qubit transformations. All other transformations from $SU(4)$ that are not elements of $SU(2)\otimes SU(2)$ are nonlocal. Following the work of \cite{makhlin}, we consider nonlocal transformations in terms of equivalence classes and call two gates $\mathcal{U}_1, \mathcal{U}_2$ locally equivalent if they differ only by applying arbitrary local operations: $\mathcal{U}_2=K_1 \mathcal{U}_1 K_2;\; K_1,K_2\in SU(2)\otimes SU(2)$. An arbitrary nonlocal transformation over a system of two qubits can be written in the form:
\BY
&&\mathcal{U}=\exp\{-i\sum\limits_{j\in{x,y,z}}\alpha_j\h\sigma^{(1)}_j\h\sigma_j^{(2)}\},\label{param}
\EY
where $\h\sigma^{(k)}_j$ are Pauli operators acting on the k-th qubit, $\alpha_j$ are real coefficients that can be found for an arbitrary unitary matrix of the two-qubit transformation.

\begin{figure}
	\includegraphics[width=8.4cm]{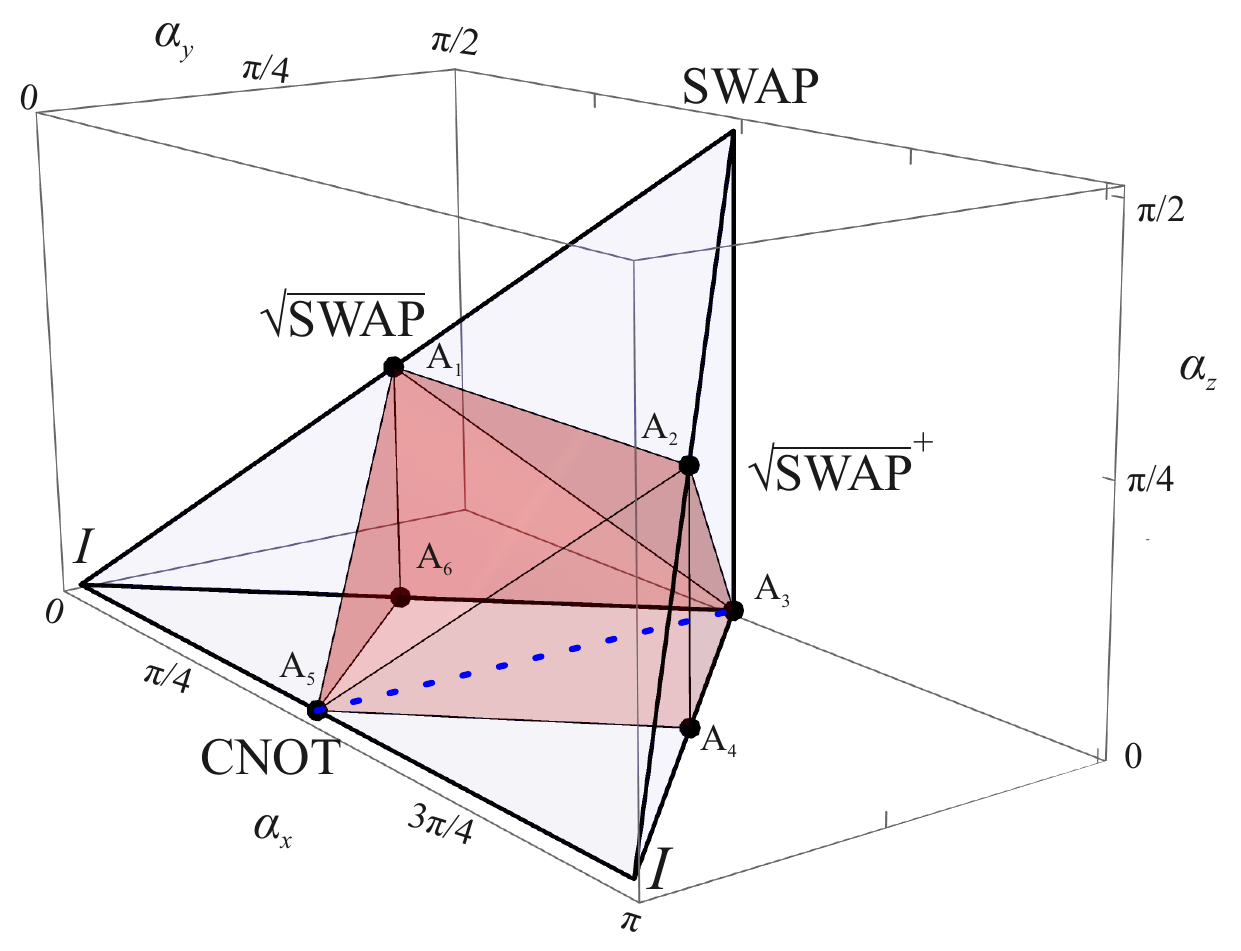}\caption{Weyl chamber for two-qubit transformations: a tetrahedron drawn in coordinates $\[\alpha_x,\alpha_y,\alpha_z\]$, each point of which corresponds to some equivalence class of nonlocal two-qubit transformations. The red polygon $A_1 A_2 A_3 A_4 A_5 A_6$ denotes the class of perfect entanglers, the blue dashed line denotes the class of special perfect entanglers. Main equivalence classes are also marked in the figure: identity equivalence class I([0,0,0] and [$\pi,0,0$]), CNOT [$\pi/2,0,0$], SWAP[$\pi/2, \pi/2, \pi/2$], $\sqrt{SWAP}[\pi/4,\pi/4,\pi/4]$, and $(\sqrt{SWAP})^\dag[3\pi/4,\pi/4,\pi/4]$) classes.}\label{figWeyl}
\end{figure}

 To find the coefficients $\alpha_j$, we diagonalise the matrix $\mathcal{U}$ by writing it in the Bell's basis:
\BY
&&M=\mathcal{Q}^\dag\mathcal{U}\mathcal{Q}\label{13},\\
&&M_{kj}=\exp\{-i\lambda_j\}\delta_{kj},\\
&&\mathcal{Q}=\frac{1}{\sqrt{2}}\begin{pmatrix}1&0&0&i\\0&i&1&0\\0&i&-1&0\\1&0&0&-i\end{pmatrix},\label{15}
\EY
 where $\delta_{kj}$ is the Kronecker symbol. The matrix $\mathcal{Q}$ is defined so that $\mathcal{Q}^\dag$ makes the rotation from computational basis of two-qubit space ($\ket{i}_1\ket{j}_2, \{i,j\}\in\{0,1\}$, lower index indicates the number of qubit) to the Bell's basis defined as follows:
\BY&&\ket{\Phi_1}=\frac{1}{\sqrt{2}}\(\ket{0}_1\ket{0}_2+\ket{1}_1\ket{1}_2\)\\
&&\ket{\Phi_2}=\frac{i}{\sqrt{2}}\(\ket{0}_1\ket{1}_2+\ket{1}_1\ket{0}_2\)\\
&&\ket{\Phi_3}=\frac{1}{\sqrt{2}}\(\ket{0}_1\ket{1}_2-\ket{1}_1\ket{0}_2\)\\
&&\ket{\Phi_4}=\frac{i}{\sqrt{2}}\(\ket{0}_1\ket{0}_2-\ket{1}_1\ket{1}_2\)
 \EY
 The eigenvalues of the  diagonal matrix $M$ could be expressed through the $\alpha_j$ as follows:
\BY
&&\lambda_1=\alpha_x-\alpha_y+\alpha_z,\;\;\lambda_2=-\alpha_x+\alpha_y+\alpha_z,\label{lam1}\\
&&\lambda_3=-\alpha_x-\alpha_y-\alpha_z,\;\;\lambda_4=\alpha_x+\alpha_y-\alpha_z.\label{lam2}\EY
In equations (\ref{lam1})--(\ref{lam2}), the global phase is omitted, which does not affect the equivalence class of the transformation. By solving the system of equations (\ref{lam1})--(\ref{lam2}), we can recover $\alpha_j$ up to the global phase. As shown in \cite{whaley, Balakrishnan}, it is convenient to define a Weyl chamber (see Fig. \ref{figWeyl}) -- a tetrahedron in the space $[\alpha_x,\alpha_y,\alpha_z]$, where different points of the tetrahedron represent different equivalence classes of nonlocal transformations. However, this representation, although it allows us to give a simple geometric interpretation, does not allow us to quantify the characteristics of nonlocal gates. In the \cite{makhlin} it was shown that it is possible to construct two quantities (one real and one complex) -- \textit{local invariants} defined by $[\alpha_x,\alpha_y,\alpha_z]$, invariant with respect to local transformations, which allow us to introduce a classification of nonlocal transformations called equivalence classes:
\BY
G_1[\mathcal{U}]&=&\frac{\hbox{Tr}^2\[m[\mathcal{U}]\]}{16 \hbox{det}\[m[\mathcal{U}]\]}=\nn\\
&&\(\prod\limits_{j\in{x,y,z}}\cos\(2\alpha_j\)-i\prod\limits_{j\in{x,y,z}}\sin\(2\alpha_j\)\)^2,\;\;\;\\
G_2[\mathcal{U}]&=&\frac{\hbox{Tr}^2\[m[\mathcal{U}]\]-\hbox{Tr}\[(m[\mathcal{U}])^2\]}{4 \hbox{det}\[m[\mathcal{U}]\]}\nn\\
&&=\sum\limits_{j\in{x,y,z}}\cos\(4\alpha_j\),\;\;\;\;\;\;
\EY
where
\BY
&&m[\mathcal{U}]=M^TM.
\EY
The first invariant is complex and the second is purely real. The values of the real and imaginary parts of the invariant $G_1$, as well as the value of $G_2$, completely determine the equivalence class of the gate: two transformations from the same class are locally equivalent and can be reduced to each other by local operations. Thus all unitary two-qubit transformations can be assigned to one or the other class, e.g. if a transformation is performed independently over only one of the two qubits and the other qubit remains unchanged, then such single-qubit local rotations have the same values of the invariants $\{1,3\}$ and belong to the equivalence class of the identical transformation. The two-cubit operation $SWAP$, although not entangling, has invariant values $\{-1,-3\}$, which does not allow us to consider this transformation local. Of particular interest to us are two large classes of states consisting of multiple equivalence classes: perfect entanglers (PE) (marked in Fig. \ref{figWeyl} with a red polygon) and special perfect entanglers (SPE) (marked with a dashed blue line) \cite{PE_SPE2}. Gates from these classes have entangling properties and are important elements for constructing quantum computation schemes in discrete variables. We first of all focus on the search for gates from PE and SPE classes. For this purpose in the next subsection we apply the above theoretical constructions to the transformation of a set of two-qubit states in the QND-rotation-QND protocol to analyze the restrictions on the equivalence classes of the transformation possible in this scheme.
\subsection{QRQ protocol unitarity and limitations on equivalence classes}
Let us now analyze the light-atomic QRQ transformation  in terms of equivalence classes of two-qubit nonlocal operations. We consider only one two-qubit subsystem, keeping in mind, however, that all transformations take place in parallel over the whole system. We can write the input separable state of the two-qubit subsystem in the following form
\BY
&&\ket{\psi}_{in}=\left(c_0\ket{0}_1+c_1\ket{1}_1\right)\otimes\left(t_0\ket{0}_2+t_1\ket{1}_2\right)=\nn\\
&&\left(c_0 \h A_0^\dag +c_1\h A_1^\dag\right)\left(t_0 \h B_0^\dag +t_1\h B_1^\dag\right)\ket{vac}.
\EY
Here, the probability amplitudes of the two-qubit state obey the standard normalization conditions: $c_0^2+c_1^2=t_0^2+t_1^2=1$. The physical states encoding the qubit states are single-excitation states, that is, for every two-qubit system there are two excitations -- one in the field and one in the atomic systems. During evolution, the field and atomic spin coherence modes with the same OAM interact -- the quadratures of the light mode are added to the quadratures of the bosonic coherence operators and vice versa, so that the states of the physical systems are correlated in phase space. But the problem of defining the type of the two-qubit transformation is rather complicated. There are several processes occurring in the two-qubit system : there is a transformation that does not take us outside the two-qubit space, and, as always in entangling bosonic transformations, there is an excitation bunching effect, and, since the Hamiltonian (\ref{hqnd}) does not preserve the number of excitations, there is a vacuum term.  Using the Bogoliubov transformation (\ref{Bog}), the unnormalized output state can be represented in the following form:
\BY
\ket{\psi}_{out}&=&\mathcal{U}_{QRQ}\ket{\psi}_{in}+f_{NQ}\ket{NQ}+\nn\\
&&(c_0t_0+c_1t_1)f_{vac}\ket{vac}\label{out},\\
\mathcal{U}_{QRQ}&=&\[\Re[f_I]\h I+i \Im[f_I]\h I+ f_S \h S\].
\EY
Here $f_I, f_S, f_{NQ}, f_{vac}$ are the non-normalised probability amplitudes, $\mathcal{U}_{QRQ}$ is the two-qubit transformation matrix in the QRQ protocol, $\h I$ is the unit transformation of two qubits, $\h S$ is the SWAP transformation matrix over the input state swapping the first and second qubits: $\h S\ket{\psi}_{in}=(t_0\ket{0}_1+t_1\ket{1}_1)\otimes(c_0\ket{0}_2+c_1\ket{1}_2)$, $\ket{vac}$ -- vacuum term, $\ket{NQ}$ -- contribution of \textit{non-two-qubit} states with bunched excitations in field and/or atomic modes:
\BY
&&f_{NQ}\ket{NQ}=f_L\left(c_0 \h A_0^\dag +c_1\h A_1^\dag\right)\left(t_0 \h A_0^\dag +t_1\h A_1^\dag\right)\ket{vac}+\nn\\
&&+f_A\left(c_0 \h B_0^\dag +c_1\h B_1^\dag\right)\left(t_0 \h B_0^\dag +t_1\h B_1^\dag\right)\ket{vac}=\nn\\
&&=f_L\ket{NQL}+f_A\ket{NQA},
\EY
where $f_A, f_L$ -- probability amplitudes, $\ket{NQA}$ -- state with two excitations bunched in the atomic medium, $\ket{NQL}$ -- state with two excitations bunched in the field system. "Non-two-qubit" means that the resulting state could not be described as a state of two qubits, in our case it is the states with two excitations bunched only in the field/atomic medium. Since we encode one qubit with only one excitation, every two-qubit state contains strictly one excitation in the field and another one in the atomic ensemble, and physical states with bunched excitations does not have a representation into two-qubit logical space.

	From the Eq. (\ref{out}) we see that the normalization of the output state depends on the input state coefficients $c_0, t_0$ and for some input states (e.g. $\ket{1}_1\ket{0}_2$) the vacuum contribution to the output state vanishes, but in the further analysis we assume that the factor $(c_0 t_0+c_1 t_1)$ takes its maximum value equal to one, i.e. in the probability analysis we estimate the upper bound of the vacuum contribution.
	
	We are interested primarily in the coefficients $f_I, f_S$, which determine the equivalence class and hence the entangling characteristics of the two-qubit gate. The vacuum and non-two-qubit terms are taken into account in probability and normalisation calculations, but we write out explicitly only the coefficients $f_I, f_S$ for the sake of simplicity:
\BY
f_I(\xi_1,\xi_2,\theta_1,\theta_2)&=&-\frac{1}{4}\left(-2i+e^{i\theta_2}\xi_1\xi_2\sin(\theta_1)\right)\times\nn\\
&&\left(-2i+e^{i\theta_1}\xi_1\xi_2\sin(\theta_2)\right),\label{fI}\\
f_S(\xi_1,\xi_2,\theta_1,\theta_2)&=&-\frac{1}{4}\left(\xi_1^2+\xi_2^2+2\xi_1\xi_2\cos(\theta_1-\theta_2)\right)\label{fS}.\;\;\;\;\;\;\;\;\;\EY
Note that the coefficient $f_S$ is purely real and depends on the difference of rotation angles of the first and second qubits. Following the logic of the previous subsection, we diagonalize the two-qubit transformation matrix $\mathcal{U}_{QRQ}$ by writing it in the Bell's basis:
\BY
&&\mathcal{Q}^\dag\mathcal{U}_{QRQ}\mathcal{Q}=\begin{pmatrix}f_I+f_S&0&0&0\\0&f_I+f_S&0&0\\0&0&f_I-f_S&0\\0&0&0&f_I+f_S\end{pmatrix}.\;\;\;\;\;\;\;\label{qrq}\EY
It should be noted that for arbitrary values of the parameters $\xi_1,\xi_2, \theta_1, \theta_2$ (see. (\ref{xi}), (\ref{Bog})) the matrix $\mathcal{U}_{QRQ}$ is not unitary, i.e. we cannot in general define an equivalence class, since the parameterization of the two-qubit transformation (\ref{param}) discussed in the previous subsection, as well as all the subsequent analysis, are only relevant for unitary matrices. Thus, it seems important to discuss in detail the restrictions imposed on the parameters by the unitarity requirement.
	
	Since, according to (\ref{13})-(\ref{15}), an arbitrary unitary matrix $\mathcal{U}$ must be diagonal in the Bell basis and have complex exponents on the diagonal, we use the Eqs. (\ref{13})-(\ref{15}) as a criterion for unitarity: the matrix $\mathcal{U}_{QRQ}$ is unitary only if all eigenvalues are complex numbers with modulus equal to one. The matrix (\ref{qrq}) can be reduced to this form by renormalisation (by taking out $|f_I+f_S|$ as the probability amplitude) only if there is a restriction on the coefficients:
\BY
|f_I+f_S|=|f_I-f_S|\Leftrightarrow\Re[f_I]f_S=0.\label{baza}
\EY
By identifying the values of parameters for which the Eq. (\ref{baza}) holds, we are able to analyze the matrix in terms of equivalence classes. By comparing (\ref{qrq}) and (\ref{lam1}), (\ref{lam2}), we can conclude that for the considered protocol the coordinates $[\alpha_x,\alpha_y,\alpha_z]$ are not independent quantities at any parameter values, since the equality of the first, second and fourth eigenvalues leads to the following condition:
\BY
\lambda_1=\lambda_2=\lambda_4\Leftrightarrow\alpha_x=\alpha_y=\alpha_z=\alpha.
\EY

 Thus, we can only perform two-qubit transformations for which all three coordinates on the Weyl cell are equal, that is, lying on lines between Identity and SWAP (see Fig. \ref{figWeyl} ), and the only intersection of these lines with the class PE, is the entangling transformation $\sqrt{\hbox{SWAP}}$ and its hermitian conjugate. An important transformation from SPE, the CNOT gate included in the universal set of operations can be realized by applying two consecutive $\sqrt{\hbox{SWAP}}$ operations with an intermediate single-qubit rotation. In the next section, we analyze several important solutions of (\ref{baza}), focusing on possible implementations of $\sqrt{\hbox{SWAP}}$ and calculating the gate probability. 
	\section{Two qubit gates and entanglement in QRQ}
	\begin{figure*}
		\includegraphics[width=18.7cm]{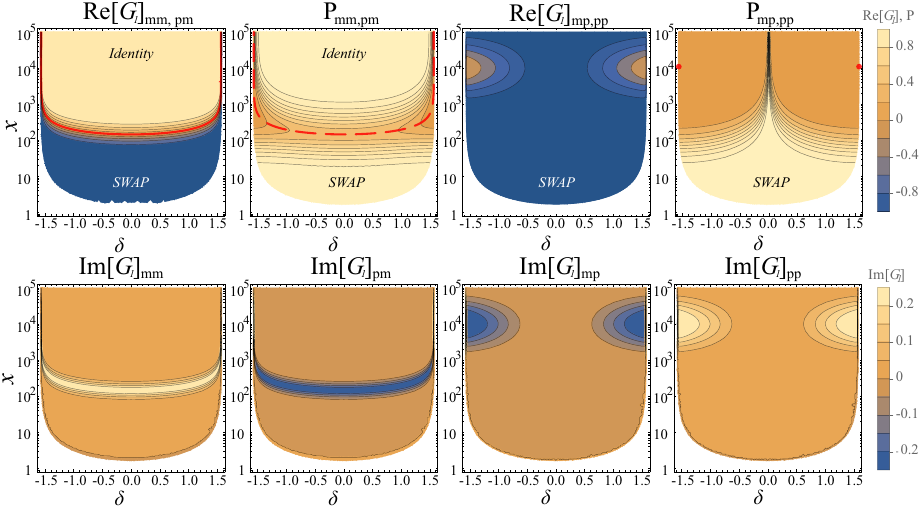}
		\caption{The real and imaginary parts of the local invariant $G_1$ and the transformation probability $P$ for different solutions of $\phi_{en}$ depending on the parameters $x,\delta$ at $\nu=100$.}\label{Fig1}
	\end{figure*}
Solutions of the Eq. (\ref{baza}) contain several interesting regimes of the QRQ protocol. As shown in Appendix A, the protocol can be reduced to non-entangling transformations when considering the special cases $\Re[f_I]=\Im[f_I]=0$ or $f_S=0$. In these special cases, the output state contains either only a two-qubit local transformation ($f_S=0$) and vacuum term, or only a SWAP transformation ($\Re[f_I]=\Im[f_I]=0$) and non-two-qubit terms. In this section we focus on analyzing the most interesting regime with respect to the generation of entanglement when the system contains both an Identity transformation with a purely imaginary coefficient and a SWAP transformation, i.e. we consider the solution of the following equation:
\BY
\Re[f_I]&=&1-\frac{1}{4}\xi_1\xi_2(4+\xi_1\xi_2\cos(\theta_1+\theta_2))\sin(\theta_1)\sin(\theta_2)\nn\\
&=&0\label{eq1}
\EY
In this case it is convenient to normalize the output state and write it in the form:
\BY
&&\ket{\psi}_{out}=\frac{\sqrt{\Im[f_I]^2+f_S^2}}{\sqrt{N}}\left(\frac{Im[f_I]}{\sqrt{\Im[f_I]^2+f_S^2}} i \h I +\right.\nn\\
&&\left.\frac{f_S}{\sqrt{\Im[f_I]^2+f_S^2}} \h S\right)\ket{\psi}_{in}+\frac{f_{NQ}}{\sqrt{N}}\ket{NQ}+\frac{f_{vac}}{\sqrt{N}}\ket{vac}\;\;\;\;\;\;\\
&&N=\Im[f_I]^2+f_S^2+f_{NQ}^2+f_{vac}^2,
\EY
where $N$ is the renormalization factor. Let us introduce the gate probability $P$ and the  the local invariants $G_1,G_2$, which for $\Re[f_I]=0$ can be written in a simplified form:
\BY
&&P=\frac{\Im[f_I]^2+f_S^2}{N}=\frac{\Im[f_I]^2+f_S^2}{\Im[f_I]^2+f_S^2+f_{NQ}^2+f_{vac}^2}\;\;\;\;\\
&&G_1=\frac{(-\Im[f_I]^2 + i \Im[f_I] f_S+ f_S^2)^2}{((\Im[f_I]- i f_S)^3 (\Im[f_I] + i f_S))}\\
&&G_2=-3+\frac{6 \Im[f_I]^2}{\Im[f_I]^2+f_S^2}
\EY

As one can see from the Eqs. for the amplitudes $f_I, f_S$ (\ref{fI}), (\ref{fS}), to simplify the analysis it is convenient to pass from the set of physical parameters $\xi_1,\xi_2, \theta_1, \theta_2$, which are the effective interaction times and qubit rotation angles, to the set $x,\nu,\phi,\delta$:
\BY
x=\xi_1 \xi_2;\;\;\nu=\xi_2;\;\;\phi=\theta_1+\theta_2;\;\;\delta=\theta_1-\theta_2
\EY
With this re-definition, each of the coefficients $f_I, f_S$ depends on only three new parameters:
$$f_I=f_I(x,\phi,\delta);\;f_S=f_S(x,\nu,\delta)$$
We solve the equation (\ref{eq1}) with respect to $\phi$, because $f_S$ does not depend on it, and we have more possibilities for further optimization of the scheme. The solution is some angle $\phi_{en}$:
\BY
\phi_{en}&=&\pm\arccos\left[-\frac{2}{x}+\frac{\cos(\delta)}{2x}\right.\nn\\
&&\left.\pm\frac{\sqrt{x^2\cos^2(\delta)+8 x\cos(\delta)-16}}{2x}\right]\label{phiS}\;\;\;\;\;\;\;\;
\EY

The index $en$ hints that such angle values allow us to realise entangling transformations. The existence of such solution naturally imposes a restriction on the parameter $x$:
\BY
x\geq\frac{4\sqrt{2}}{|\cos(\delta)|}-\frac{4}{\cos(\delta)}
\EY

We classify the four possible sets of solutions as $\{pp,pm,mp,mm\}$ according to the first and second signs in (\ref{phiS}). Analyzing all possible output states for all solutions $\phi_{en}$ in explicit form is rather cumbersome, so we focus on the numerical computation of the probability $P$ and the values of the local invariants $G_1, G_2$. From Eqs. (41)-(42) one can see that the values of the angle $\phi_{en}$ and the constraints on the parameter $x$ depend also on the sign of $\cos\delta$, so we separately consider the cases $\delta\in(-\pi/2,\pi/2)$ and $\delta\in(\pi/2,3\pi/2)$.

By restricting ourselves to the set of solutions (\ref{phiS}) we have three governing parameters. Let us first observe how the difference in the rotation angles of the two qubits affects the equivalence class of the transformation by fixing the parameter $\nu=100$.  In Fig. \ref{Fig1} we present the dependencies of the real and imaginary parts of the invariant $G_1$ on $\delta\in(-\pi/2,\pi/2)$ and $x\in[(4\sqrt{2}-4)/\cos{(\delta)},10^5]$.  The regions where $\Re[G_1]=1$ correspond to the class of Identity  transformation, $\Re[G_1]=-1$ correspond to the class of SWAP transformation. The region in the middle, where the local invariant $G_1$ becomes purely imaginary and takes the value $\pm i/4$, i.e., $\Re[G_1]=0$, is the most interesting for us -- these locus of points are marked by red solid and dashed lines. This invariant values corresponds to the operations locally equivalent to $\sqrt{SWAP}$ (for $\Im[G_1]=+i/4$) or $(\sqrt{SWAP})^\dag$ (for $\Im[G_1]=-i/4$), which are entangling and belong to the Perfect Entangler (PE) \cite{PE_SPE, PE_SPE2} class. For $mm,pm$ solutions PE operations can be performed at any value of $\delta$, while for the other two solutions -- only at the points $\delta=\pm\pi/2,x=\nu^2$. It should be noted that in the general case for full identification of the transformation in terms of equivalence classes it is also necessary to know the value of the invariant $G_2$, but in the considered situation the dependence of $G_2$ on the parameters in general repeats the one for $Re[G_1]$, and fully confirms the above.
 	
 	It is important to note that all intermediate transformations for which $\Re[G_1]\in(-1,1)$ are also entangling to some extent, but none of them belong to the Perfect Entanglers class, so we do not focus on them, keeping in mind, however, that they may also be interesting for the implementation of quantum algorithms. 
 	
 	 Fig. \ref{Fig1} shows the probabilities $P$, where different types of transformations defined by the values of the local variant $G_1$ are highlighted with captions and colour. 	 It can be seen that the highest PE transformation probabilities are achieved in the region $\delta=0$ for $mm,pm$ solutions. For the solutions $mp,pp$, however, the points of PE operations do not lie in the region of high probabilities, moreover, the constraint on the parameters (42) is no longer meaningful for angles $\phi_{en}$ close to $\pi/2$, so we exclude them from the analysis. 

\begin{figure}
	\includegraphics[width=8.3cm]{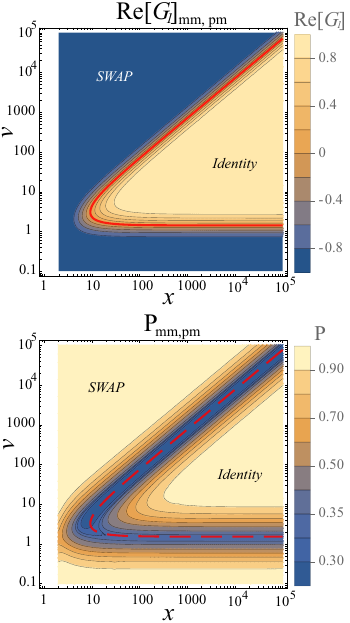}
	\caption{The real part of the local invariant $G_1$ for $mm, pm$ solutions and the transformation probability $P$ as a function of the parameters $x,\nu$ when $\delta=0$.}\label{Fig2}
\end{figure}

 Identifying the most interesting value of the angle $\delta=0$ (which corresponds to the identical qubit rotation angles $\theta_1=\theta_2$), we follow the changes in the probability and value of the $G_1$ invariant only on the effective interaction times given by the parameters $x$ and $\nu$. We are focused on the $mm$ solution, keeping in mind that the $pm$ solution is similar up to the $G_1$ complex conjugation. In Fig. \ref{Fig2} we can see that there are two stable regimes of the entangling operation (locus of points with $\Re[G_1]=0$, marked by the bold red line), which, however, differ in probabilities -- the sloping line has probability of $\sqrt{SWAP}$ tending to $1/4$ at $\nu\rightarrow\infty$, and not exceeding $0.3$ for the small $\nu$. The horizontal line corresponds to the value $\nu=\sqrt{2}$ and the probability reaches $1/3$ in a wide range of the parameter $x$, which without loss of generality can be taken to be much larger than one. Calculating the value of $\phi_{en}$ according to the Eq. (\ref{phiS}) it is possible to write down the conditions for performing PE operations through the values of physical parameters (see Table 1 ).
\begin{table}[h]
	\caption{Values of parameters for PE operations for $\delta\in(-\pi/2,\pi/2)$}
	\centering\begin{tabular}{|c|c|c|c|}
	\hline
Computational param.&Physical param.&Equiv. Class&Prob.\\
	\hline
$ \begin{array}{@{}l@{}}x\gg1\\
		\nu=\sqrt{2}\\
		\delta=0\\
		\phi=\pi/2\end{array}$ &$ \begin{array}{@{}l@{}}\xi_1\gg1\\
		\xi_2=\sqrt{2}\\
		\theta_1=\pi/4\\
		\theta_1=\pi/4\end{array}$&$\sqrt{SWAP}$&$\displaystyle\frac{1}{3} $ \\
	\hline
	 $ \begin{array}{@{}l@{}}x=\sqrt{2}\nu\\
			\nu\gg1\\
			\delta=0\\
			\phi=\pi/2\end{array}$ &$\begin{array}{@{}l@{}}
			\xi_1=\sqrt{2}\\
			\xi_2\gg1\\
			\theta_1=\pi/4\\
			\theta_2=\pi/4\\
			\end{array}$&$\sqrt{SWAP}$&$\displaystyle\frac{1}{4} $ \\
	\hline
	$ \begin{array}{@{}l@{}}x\gg1\\
		\nu=\sqrt{2}\\
		\delta=0\\
		\phi=-\pi/2\end{array}$ &$ \begin{array}{@{}l@{}}\xi_1\gg1\\
		\xi_2=\sqrt{2}\\
		\theta_1=-\pi/4\\
		\theta_1=-\pi/4\end{array}$&$(\sqrt{SWAP})^\dag$&$\displaystyle\frac{1}{3} $ \\
	\hline
	 $ \begin{array}{@{}l@{}}x=\sqrt{2}\nu\\
		\nu\gg1\\
		\delta=0\\
		\phi=-\pi/2\end{array}$ &$\begin{array}{@{}l@{}}
		\xi_1=\sqrt{2}\\
		\xi_2\gg1\\
		\theta_1=-\pi/4\\
		\theta_2=-\pi/4\\
	\end{array}$&$(\sqrt{SWAP})^\dag$&$\displaystyle\frac{1}{4} $ \\
	\hline
\end{tabular}
\end{table}
 
\begin{figure*}
	\includegraphics[width=18.7cm]{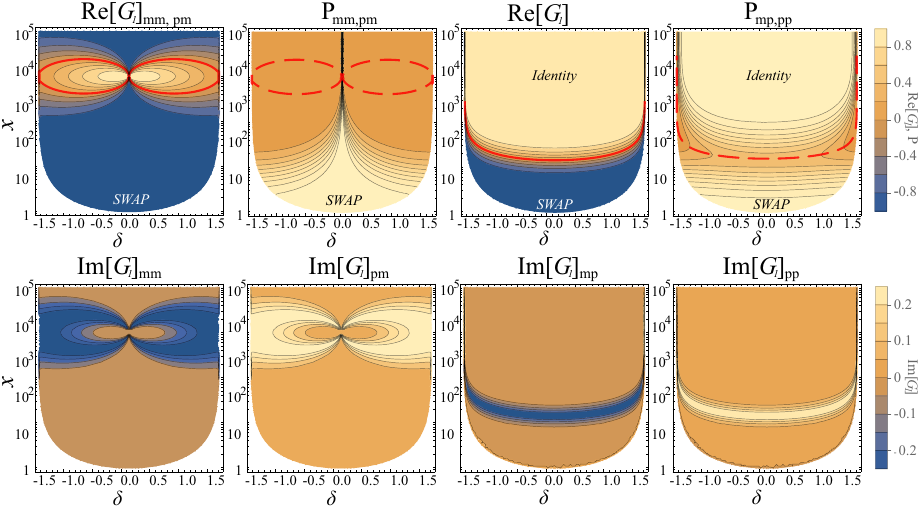}
	\caption{The real and imaginary parts of the local invariant $G_1$ and the transformation probability $P$ for different solutions of $\phi_{en}$ depending on the parameters $x,\delta$ at $\nu=100$.}\label{Fig3}
\end{figure*}
 In general, the condition for the PE operation can be written in the form \BY|\Im[f_I(x,\phi_{en},\delta)]|=|f_S(x,\nu,\delta)|\label{PECond}\EY and we are not obliged to pick values of constants presented in Table 1 -- we can use all sets of values which ensure the zero of the $\Re[G_1]$, however, the values from Table 1 provide the highest probabilities of transformations. The analytical solution of the Eq. (\ref{PECond}) is rather cumbersome, so we do not give it in the text, limiting ourselves to the analysis of the asymptotic behavior.

Table 1 shows that at least one of the interaction constants, $\xi_1$ or $\xi_2$, must necessarily be large to ensure entangling operation. The increase of these constants can be provided by increasing the effective interaction time of the driving field impulse with the atomic ensemble, or by increasing the atomic density. 
As can be seen from Table. 1, the equivalence class of the resulting transformation in the considered regime does not depend on the order of application of QND operations, i.e., it makes no difference which of the constants $\xi_1 $ or $\xi_2 $ we choose to be equal to $\sqrt{2}$ and which one we take to infinity, but the probabilities vary significantly. If the first QND operation occurs with a large interaction constant (which can be interpreted as large interaction time), then before the rotation of the atomic and light systems, a notable correlation between them is already established. So we perform in fact not two independent rotations, but a complicated phase transformation of the light and atomic variables, after which the value of the second interaction constant $\xi_2=\sqrt{2}$ provides the desired interference of states. In the other case, when the values of the constants have chosen in reverse, after the first short interaction the systems appear to be correlated relatively weakly. The interference of states necessary for the PE operation can be achieved only after the large time of the second QND interaction.

 Let us proceed to the analysis of the system in the range $\delta\in(\pi/2,3\pi/2)$. Fig. \ref{Fig3} shows the dependencies of the real and imaginary parts of the invariant $G_1$ on $\delta\in(\pi/2,3\pi/2)$ and $x\in[(4\sqrt{2}+4)/\cos{(\delta)},10^5]$ with the constant value of $\nu=100$. It can be observed that the dependencies for the solutions $pp,mp$ in this range of angles repeat those for $mm,pm$ in the range $\delta\in(-\pi/2,\pi/2)$ presented in Fig. \ref{Fig1}. The values of the parameters for the PE transformation are similar to those presented in Table 1 up to the shift of $\delta$ by $\pi$, and hence the values of the angles $\theta_1,\theta_2$ by $\pm3\pi/4, \mp\pi/4$, respectively. For the solutions $mm,pm$ we also observe similarity with the previously considered range of $\delta$, and the PE operation's locus of points do not coincide with the regions of high probabilities. However, the behavior of local invariants and probabilities in the vicinity of the point $\delta=\pi, x=\nu^2$ (where the red curves intersect) in this case requires a more detailed consideration.
 	 \begin{figure}
 		\includegraphics[width=8.3cm]{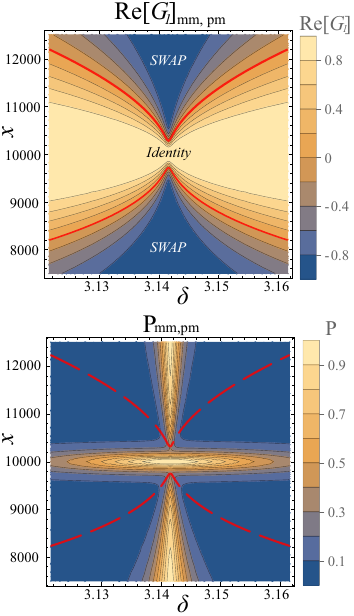}
 		\caption{The real and imaginary parts of the local invariant $G_1$ and the transformation probability P for solutions $pm,mm$ (enlarged region of Fig. \ref{Fig3} in the region around the point $\delta=\pi$) as a function of the parameters $x,\delta$ at $\nu=100$.}\label{Fig4}
 	\end{figure}
 	
 	Fig. \ref{Fig4} shows the real part of $G_1$ and the probability $P$ in the region of the point of interest. We see that at $\delta=\pi,x=\nu^2$ a deterministic Identity transformation is performed, whereas if we move away from this point along the $x$-axis, independently of the direction, we'll get into the region of deterministic $SWAP$ operation. The red lines, as before, mark the zeros of $G_1$ -- the locus of points of the PE operations $\sqrt{SWAP}$ or $(\sqrt{SWAP})^\dag$ (depending on the sign chosen in (37)). When $\delta=\pi$, these lines fall into the regions of high probability.
 	 	 \begin{figure}
 		\includegraphics[width=8.2cm]{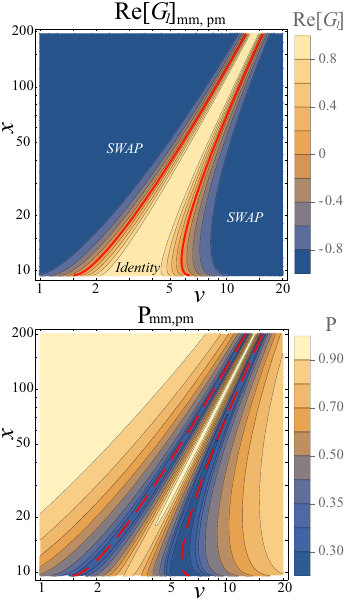}
 		\caption{Real part of the local invariant $G1$ and probability of transformation $P$ for solutions $pm,mm$  as  functions of the parameters $x,\nu$ for $\delta=\pi$.}\label{Fig5}
 	\end{figure}
 	
 	  \begin{table}[h]
 		\caption{Parameter values for performing PE operations for $\delta\in(\pi/2,3\pi/2)$ for $mm,pm$ solutions. The signs "$+,-$" after the probability indicate from which direction the probability tends to the specified value}
 		\centering\begin{tabular}{|c|c|c|c|}
 			\hline
 			Computational param.&Physical param&Equiv. Class&Prob.\\
 			\hline
 			$ \begin{array}{@{}l@{}}
 				x\gg1\\
 				\nu=\sqrt{x}+\sqrt{2}\\
 				\delta=\pi\\
 				\phi=\pi\\
 			\end{array}$ &$ \begin{array}{@{}l@{}}
 				\xi_1+\sqrt{2}\gg1\\
 				\xi_2-\sqrt{2}\gg1\\
 				\theta_1=\pi\\
 				\theta_2=0\\
 			\end{array}$&$\sqrt{SWAP}$&$\displaystyle\frac{4}{13}- $ \\
 			\hline
 			$ \begin{array}{@{}l@{}}
 				x\gg1\\
 				\nu=\sqrt{x}-\sqrt{2}\\
 				\delta=\pi\\
 				\phi=\pi\\
 			\end{array}$ &$ \begin{array}{@{}l@{}}
 				\xi_1-\sqrt{2}\gg1\\
 				\xi_2+\sqrt{2}\gg1\\
 				\theta_1=\pi\\
 				\theta_2=0\\
 			\end{array}$&$\sqrt{SWAP}$&$\displaystyle\frac{4}{13}+ $ \\
 			\hline
 			$ \begin{array}{@{}l@{}}
 				x\gg1\\
 				\nu=\sqrt{x}+\sqrt{2}\\
 				\delta=\pi\\
 				\phi=-\pi\\
 			\end{array}$ &$ \begin{array}{@{}l@{}}
 				\xi_1+\sqrt{2}\gg1\\
 				\xi_2-\sqrt{2}\gg1\\
 				\theta_1=0\\
 				\theta_2=-\pi\\
 			\end{array}$&$(\sqrt{SWAP})^\dag$&$\displaystyle\frac{4}{13}- $ \\
 			\hline
 			$ \begin{array}{@{}l@{}}
 				x\gg1\\
 				\nu=\sqrt{x}-\sqrt{2}\\
 				\delta=\pi\\
 				\phi=-\pi\\
 			\end{array}$ &$ \begin{array}{@{}l@{}}
 				\xi_1-\sqrt{2}\gg1\\
 				\xi_2+\sqrt{2}\gg1\\
 				\theta_1=0\\
 				\theta_2=-\pi\\
 			\end{array}$&$(\sqrt{SWAP})^\dag$&$\displaystyle\frac{4}{13}+ $ \\
 			\hline
 		\end{tabular}
 	\end{table}
 	
 	To follow the probability value, we study its dependence on the parameters $x, \nu$ at $\delta=\pi$. In Fig. \ref{Fig5} we can see that at each value of $x$ there are two values of $\nu$ symmetric with respect to $\nu^2$, providing a PE transformation. In this case, the probabilities of the first and second regimes are different -- the probability in the region of low $\nu$ is higher than in the region of high values of $\nu$, but at high values of $x$ both probabilities asymptotically tend to $4/13$.
 	  
 	Such asymmetry in the probabilities at small interaction constants is caused by the constraints on $x$ from below, dictated by the requirement of the angle $\phi_{en}$ to be real. At large interaction constants, we can write the conditions for the PE operation through the physical values of the parameters explicitly (see Table 2). We need to make both interaction constants large simultaneously under the interference conditions: if we make the constants equal, we reach the region of the Identity transformation; to perform the PE operation we need to ensure that the two constants differ by $\pm2\sqrt{2}$.
 	
  To summarize, the output state of the system for all cases satisfying (\ref{PECond}) can be written as (we omit the phase factors in the notation, and also keep in mind that the operation $\sqrt{SWAP}$ in the expressions below can be replaced by $(\sqrt{SWAP})^\dag$):
 \BY
&& \ket{\psi}_{out}=\sqrt{P}\sqrt{SWAP}\ket{\psi}_{in}+\sqrt{P}\left(\frac{\ket{NQL}+\ket{NQA}}{\sqrt{2}}\right)\nn\\
&&+\sqrt{(1-2P)}\ket{vac},\\
 &&P=P(\xi_1,\xi_2,\theta_1,\theta_2)\in\[1/4,1/3\]
 \EY

 Thus, in all cases when PE transformation occurs with high probability, in addition to two-qubit entanglement, an equal-weighted superposition of states with bunching of excitations (a state caused by the Hong-Ou-Mandel effect) in the atomic ($\ket{NQA}$) and in the light ($\ket{NQL}$) media is generated. There are also parameter values that lead to bunching of excitations only in the light or atomic media, such regimes are described in detail in Appendix A.
 
  Summarising all above, we can formulate following steps to perform the PE transformation over the ensemble of two-qubit subsystems: 
\begin{steps}
	\item \textit{Preparation of the initial quantum state of light.} To initialise the whole quantum state of light, which contains K/2 qubits, one should prepare each qubit. One qubit with number $j$ could be prepared by generation of a single photon in the superposition state of a two OAM modes (for experimental technique see \cite{crypt1,SPOAM}), chosen in correspondence with Eq. (10). The whole state of light consists of K/2 photons. 
\item \textit{Prepation of the initial quantum state of atomic ensemble.} To prepare the atomic state one could  perform the step one and prepare the new system of light qubits (separated from the system, prepared on the previous step) in desired states (see Eq. (11)) firstly. To transmit the state to the atomic ensemble one could perform the quantum memory write-in protocol \cite{muzhik}. 
\item \textit{The first QND operation} On this step one need to execute the QND interaction: two pulses of light of the same duration (quantum and driving) propagate through an atomic cell (see Fig.1). After the scattering of the fields, spin coherence between lower levels is generated. In the Holstein-Primakoff approximation, one can describe spin coherence evolution in terms of harmonic oscillator, and operate with the states with single bosonic excitation of the spin coherence OAM modes.  

Amplitude and duration of a driving field pulse defines the interaction constant $\xi$ in Eq. (\ref{xi}). Values of $\xi$ to perform entangling and nonlocal operations could be found in Tables 1 and 2. The transverse area of the driving beam should be chosen to be much larger than that of quantum field and atomic cell to provide equal geometrical factors $\chi_{m-k}$ for each pair of OAM modes. 
\item \textit{Rotation of the qubits.} The rotation by an arbitrary angle $\theta_1$ in the light system can be described as an additional phase of the light beam, which can be added by using phase plates or a delay line, the rotation of atomic variables can be performed by applying an external magnetic field along the $x$-axis \cite{holog}.
\item \textit{The second QND operation} On the last step one provide another one QND interaction to fulfil the interference conditions.
\end{steps}

After the gate's operation we can perform the tomography of light OAM qubits and also measure the atomic system after the transmission the atomic quantum state back to light by Step 2. 

It should be noted that in the experiment we have many factors that affect the fidelity and probability of gates. Errors caused by imperfect implementation of the protocol are apparently the same as for quantum memory due to the similarity of models. Here we briefly discuss the main factors affecting the errors. 

Firstly, errors in the two-qubit transformation are expected to arise due to the thermal motion of atoms. Consideration of the thermal motion for our case in a quite  similar quantum memory system is presented, for example, in \cite{Tikh}. It is shown that optimisation of the field profiles allows to keep high values of quantum memory efficiency at cooling temperatures of MOT (about hundreds of $\mu K$ at storage times on the order of tens or hundreds of $\mu s$),the storage of OAM modes in EIT protocol \cite{Veiss} with similar storage times and temperatures is also experimentally demonstrated .

Since we operate with spatial modes of collective spin coherence, each element of the atomic ensemble acquires some phase as a result of interaction with light, and the transverse phase profile of the collective coherence determines the spatial OAM mode. Transverse thermal motion will lead to mixing of phase information in each spatial mode, which can be described as the presence of crosstalk between modes with neighbouring indices (a similar mechanism is described in \cite{Atm}). The longitudinal motion of atoms in our case will have a weaker effect on the quality of the conversion. This feature of QND memory has been used to demonstrate the quantum memory at room temperature with high fidelity \cite{Jul1, Jul2}. Authors used long pulses about $ms$ in the write/read stage to suppress the effect of atomic motion, but such an implementation needs large delay lines when considering multipass protocols. Given the above, requirements for atomic traps can be formulated: trap should provide good cooling primarily for the transverse degrees of freedom (e.g., \cite{MOT}).  However, if we consider only the two-qubit gate protocol without memory, the write time in the QND protocol (i.e., the gate execution time), which is in practice a microsecond or less, is an orders of magnitude shorter than the storage time, so the atomic temperature requirements can be relaxed significantly and one should expect good fidelity values of the gate at MOT temperatures.

The second important source of errors in the considered protocol is the imperfect control of the effective interaction constants (determined by the duration of the driving field pulses). As shown in Figure 6, the execution of the $\sqrt{SWAP}$ entangling operation requires the precision in the interaction constant, in contrast to SWAP and local operations. A $4\%$ deviation of $\xi_1 \xi_2$ from the required value will result in the implementation of a non-entangling local operation instead of entangling one.

Other effects that distinguish realistic systems from the theoretical model, such as inhomogeneous broadening of atomic transitions, decay to other levels, and so on, contribute less to the quality of the transformation; a review of their influence on errors in various protocols is presented in \cite{QI}. 
\section{Conclusion}
In this work we investigated the quantum non-demolition interaction of an ensemble of cold four-level atoms with a multimode field with OAM. In the case when the OAM of the driving field equals to zero, the interaction can be reduced to a pairwise interaction between field modes and spin coherence modes with the same OAM values. In such a regime, we investigated the QND-rotation-QND protocol in terms of discrete variables. The considered interaction of field and atomic states with a certain OAM allows us to identify a set of independent two-qubit subsystems when encoding logical qubit states via physical states with single excitation. By performing quadrature rotations and controlling the effective interaction constants $\xi_1,\xi_2$ through the duration and Rabi frequency of the driving field pulse, as well as the atomic density, we can perform a wide range of parallel two-qubit operations on an ensemble of two-qubit subsystems in the QRQ protocol.

To classify the entangling properties of transformations, we have considered the formalism of equivalence classes. The calculation of local invariants showed that the most entangling of all possible operations in QRQ would be $\sqrt{SWAP}$ and $(\sqrt{SWAP})^\dag$, belonging to the PE class. For the illustrative representation of the entangling properties of these gates, one can calculate the entangling power of the gate through the value of local invariant $G_1$. As shown in \cite{Balakrishnan}, entangling power is linear function of the $G_1$ modulus, so for the $\sqrt{SWAP}$ and $(\sqrt{SWAP})^\dag$ have entangling power equal to 1/6, whereas SPE gates such as CNOT have maximum possible power value equal to 2/9.  It is worth noting that, as shown in \cite{XY}, the CNOT operation included in the universal set can be decomposed as a sequence of two $\sqrt{SWAP}$ with a single-qubit operation in the middle, so in our case its execution requires four sequential QND interactions with properly chosen rotations between them.  The protocol also implements other entangling operations, that are not PE, but are of hypothetical interest in the application of particular algorithms.The choice of rotation angles of one and the other qubits allows the deterministic SWAP operation, the bunching of excitations, and the probabilistic generation of the mentioned entangling operations $\sqrt{SWAP}$ and $(\sqrt{SWAP})^\dag$. The optimization of the interaction constants allowed us to identify several regimes of entanglement generation with the highest probability of $1/3$, allowing the CNOT operation to be performed with a probability of $1/9$.

Speaking of probabilistic two-qubit gates, it is necessary to mention the heralding mechanism. Since the generation of a two-qubit entanglement is accompanied by the generation of a state with bunched excitations, we can implement a heralding mechanism similar to the one used in \cite{OBrien}, where the presence of coincidences between the photodetection acts in some channels at the output of the scheme meant that the gate was successfully operated. We can also select only those outcomes in which there is one excitation in the field medium and one excitation in the atomic medium. So we proposes the protocol with the same CNOT probability as in \cite{OBrien}, but with some advantages. Presented method allow to perform many two-qubit gates in parallel, and also to switch from one class of two-qubit operation to another.

The multidimensional nature of the chosen degree of freedom  allows us to perform many two-qubit operations simultaneously. This property of the system seems particularly important in light of recent works on a quantum processor \cite{Lukin1, Lukin2}, where it is necessary to create multi-dimensional entangled states and implement parallel operations. But there is an issue of initializing the input state of many two-qubit subsystems: for $K/2$ two-qubit subsystems we need $K$ of excitations in OAM superposition states. The logical step to reduce the number of excitations at the cost of increasing the superposition's dimensionality is the transition to qudit logic, which, however, requires additional studies of local invariants and equivalence classes.

	\section{Acknowledgements}
	
	This work was financially supported by the Theoretical Physics and Mathematics Advancement Foundation "BASIS" (grant No. 22-1-4-20-1). T.Y.G. acknowledges support by the Ministry of Science and Higher Education of the Russian Federation on the basis of the FSAEIHE SUSU (NRU)
	(Agreement No. 075-15-2022-1116).
	\appendix
		\section{Local and non-entangling transformations in QRQ protocol}
\begin{figure}
		\includegraphics[width=8.3cm]{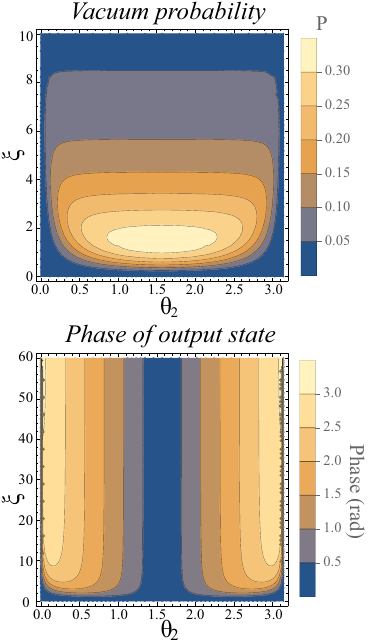}
		\caption{Top: probability of the vacuum state depending on the parameters $\xi, \theta_2$ Bottom: phase factor of the output two-qubit state as a function of the rotation angle of the second qubit and the interaction constant. Both figures are calculated under the conditions $\xi_1=\xi_2=\xi, \theta_2=\theta_1+\pi$.}\label{FigIdent}
	\end{figure}

In this section we focus on cases where the resulting two-qubit is local, or non-local but not entangling, and impose a constraint (\ref{baza}) on the parameters. One of the possible interesting cases is the transformation in which the amplitude of $f_S$ is zero and we have the gate from the Identity transformation class at the output:
	\BY
&&	f_S=-\frac{1}{4}\left(\xi_1^2+\xi_2^2+2\xi_1\xi_2\cos(\theta_1-\theta_2)\right)=0\Leftrightarrow\nn\\ &&\left\{\begin{array}{@{}l@{}}
		\xi_1=\xi_2=\xi\\
		\theta_1-\theta_2=\pi
	\end{array}\right.\,.
	\EY
	This equation can be solved with real parameters in the unique way: $\xi_1=\xi_2, \theta_1-\theta_2=\pi$ for $\theta_1,\theta_2\in\[-\pi/2,3\pi/2\]$. In this case, the unnormalized matrix of the two-qubit transformation can be written in the form (see (27), (29),(30)):
	\BY
	\mathcal{U}_{QRQ}(\xi,\xi,\theta_2+\pi,\theta_2)= \left((e^{2 i\theta_2}-1)\frac{\xi^2}{4}-1\right)^2\h I
	\EY
	The contribution of the excitation bunching terms is zero in this case, but there are the contribution of the vacuum term:
	\BY
	&&\ket{\psi}_{out}=f_I\ket{\psi}_{in}+f_{vac}\ket{vac}=\nn\\
	&&\left((e^{2 i\theta_2}-1)\frac{\xi^2}{4}-1\right)^2\ket{\psi}_{in}+\nn\\&&\frac{i}{2}\left(e^{2 i\theta_2}-1\right)\xi\left((e^{2 i\theta_2}-1)\frac{\xi^2}{4}-1\right)\ket{vac}
	\EY
	The normalization of the output state can be recovered in a standard way, then it is convenient to write the normalized state in the following form:
	\BY
	&&\ket{\psi}_{out}=\frac{|f_I|}{\sqrt{|f_I|^2+|f_{vac}|^2}}\left(\frac{f_I}{|f_I|}\ket{\psi}_{in}\right)+\nn\\&&\frac{f_{vac}}{\sqrt{|f_I|^2+|f_{vac}|^2}}\ket{vac}
	\EY
	From the expression above we see that in the considered example the QRQ protocol is reduced to multiplication of the input state by the phase factor $\displaystyle\frac{f_I}{|f_I|}$ with the probability amplitude $\displaystyle\frac{|f_I|}{\sqrt{|f_I|^2+|f_{vac}|^2}}$ (see Fig. \ref{FigIdent}). One can notice that with the growth of the parameter $\xi$, i.e. at large integral interaction times (see (\ref{xi})), the vacuum term vanishes and only the phase transformation remains. The maximum probability of the vacuum state reaches 1/3 at $\xi=\sqrt{2}, \theta_2=\pi/2$, the phase of the input state at such parameters does not change. One can see that the state transformation does not actually occur, but the vacuum term appears, i.e., energy is simply lost. This effect could be called destructive state interference in the QRQ protocol. The regimes considered in the main text then can be called constructive interference, where the appearance of the vacuum term with probability $1/3$ was also accompanied by the generation of entanglement.
		We calculated the values of the local invariants for such a QRQ transformation regime and obtained values $\{G_1, G_2\}=\{1,3\}$, which corresponds to operations from the equivalence class of the Identity transformation, i.e. in such a regime the realised operation is local and does not generate an entanglement.  
	
	The second type of solutions (\ref{baza}) is a wide set of parameters for which $\Re[f_I]=0$. The general analysis of this regime is presented in Section II of the main text, and here we focus only on the special case when both real and imaginary parts of $f_I$ are zero, i.e. we expect the realization of a transformation from the SWAP class:
		\BY
	f_I&=&-\frac{1}{4}\left(-2i+e^{i\theta_2}\xi_1\xi_2\sin(\theta_1)\right)\left(-2i+e^{i\theta_1}\xi_1\xi_2\sin(\theta_2)\right)\nn\\
	&=&0
	\EY
Eq. (A5) has two sets of solutions, different in terms of physical realization:
	\BY
	&&\left\{\begin{array}{@{}l@{}}
		\theta_1=\pm\pi/2\\
		\theta_2=\pm\arcsin\frac{2}{\xi_1\xi_2}
	\end{array}\right.;
	\left\{\begin{array}{@{}l@{}}
		\theta_1=\pm\arcsin\frac{2}{\xi_1\xi_2}\\
		\theta_2=\pm\pi/2
	\end{array}\right.;\EY
		The output state for both sets of solutions contains a term describing the SWAP transformation over the input state, an excitation bunching term, and a vacuum term, and can be represented as:
		\BY
	&&\ket{\psi}_{out}=\frac{1}{\sqrt{N}}\left( f_S \h S\ket{\psi}_{in}+f_{NQ}\ket{NQ}+f_{vac}\ket{vac}\right)\;\;\;\;\;\;\;\\
	&&\frac{f_S}{\sqrt{N}}=-\sqrt{\frac{4+\xi_1^2+\xi_2^2}{\xi_1^2+\xi^2_2+2\xi_1^2\xi_2^2}}\\
	&&\frac{f_{NQ}}{\sqrt{N}}=\frac{\sqrt{\xi_1^2\xi_2^2-4}(-i\sqrt{\xi_1^2\xi_2^2-4}\pm(\xi_2^2+2))}{\xi_2\sqrt{(4+\xi_1^2+\xi_2^2)(\xi_1^2+\xi^2_2+2\xi_1^2\xi_2^2)}}\\
	&&\frac{f_{vac}}{\sqrt{N}}=\frac{\xi_2(2i +i\xi_1^2+\sqrt{\xi_1^2\xi_2^2-4})}{\sqrt{(4+\xi_1^2+\xi_2^2)(\xi_1^2+\xi^2_2+2\xi_1^2\xi_2^2)}}
	\EY
	
	Such solutions impose conditions on the product of the interaction constants $\xi_1\xi_2\geq2$. It is interesting that the two sets of solutions differ in type of physical system, where the excitations are bunched, in the atomic medium or in the light:
	
	\BY
	&&\left\{\begin{array}{@{}l@{}}
		\theta_1=\pm\pi/2\\
		\theta_2=\pm\arcsin\frac{2}{\xi_1\xi_2}
	\end{array}\right.\Rightarrow f_{NQ}=f_L,f_A=0\\
	&&\left\{\begin{array}{@{}l@{}}
		\theta_1=\pm\arcsin\frac{2}{\xi_1\xi_2}\\
		\theta_2=\pm\pi/2
	\end{array}\right.\Rightarrow f_{NQ}=f_A,f_L=0
	\EY
 \begin{table}
	\caption{Parameter values for performing various non-entangling transformations}
	\centering\begin{tabular}{|c|c|c|}
		\hline
		Physical param&Output state&Prob.\\
		\hline
		$ \begin{array}{@{}l@{}}
			\xi_1\xi_2=2\gg1\\
			\xi_2\gg1\\
			\theta_1=\pm\pi/2\\
			\theta_2=\theta_1\\
		\end{array}$&$SWAP\ket{\psi}_{in}$&$1$ \\
		\hline
		$ \begin{array}{@{}l@{}}
			\xi_1\gg1\\
			\xi_2=\xi_1\\
			\theta_1=0\\
			\theta_2=\pm\pi/2\\
		\end{array}$&$\displaystyle\frac{(1\pm i)}{\sqrt{2}}\ket{NQL}$&$\displaystyle\frac{1}{2}$ \\
		\hline
		$ \begin{array}{@{}l@{}}
			\xi_1\gg1\\
			\xi_2=\xi_1\\
			\theta_1=\pm\pi/2\\
			\theta_2=0\\
		\end{array}$&$\displaystyle\frac{(1\pm i)}{\sqrt{2}}\ket{NQA}$&$\displaystyle\frac{1}{2}$ \\
		\hline
	\end{tabular}
\end{table}

A special case of (A6) is the choice of constants $\xi_1\xi_2=2$, where the amplitude of the bunched terms becomes zero and the QRQ transformation can be reduced to a SWAP operation with the addition of a vacuum term that vanishes when one of the constants tends to infinity, preserving $\xi_1\xi_2=2$. This regime was described in detail in \cite{BVG}. If we choose both constants to be sufficiently large (we can, for convenience, put $\xi_1=\xi_2$), then the normalized probability amplitude $f_S$  tends to zero and the output state contains the state with excitation bunching in one or another medium (depending on the chosen angles) and the vacuum term (see Table 3).

Excitations are bunched only in one of the physical systems, and in the regimes considered here we are not obtain at the output the generation of a state similar to the second term in (44), which emphasizes the connection of this quantum state with quantum interference and entanglement generation.

\end{document}